\documentclass[reprint,superscriptaddress,amsmath,amssymb,aps]{revtex4-2}

\usepackage{graphicx}
\usepackage{dcolumn}
\usepackage{bm}
\usepackage{braket}
\usepackage{mathrsfs}
\usepackage{hyperref,verbatim}
\hypersetup{
    colorlinks,%
    citecolor=blue,%
    filecolor=blue,%
    linkcolor=blue,%
   	urlcolor=blue,
   	linktoc=page
}
\allowdisplaybreaks

\usepackage{physics}
\usepackage{accents}
\usepackage{pict2e}
\usepackage[normalem]{ulem}
\usepackage{float}
\usepackage{mathtools}
\usepackage{mathrsfs}
\usepackage{setspace}
\usepackage{tikz}
\usetikzlibrary{calc}
\usetikzlibrary{decorations.pathreplacing}
\usetikzlibrary{decorations.pathmorphing}
\usetikzlibrary{decorations.markings}
\usetikzlibrary{arrows.meta}
\usetikzlibrary{positioning}

\usepackage{mathtools}
\usepackage[T1]{fontenc}
\usepackage{tipa}
\usepackage{tipx}
\usepackage{wasysym}

\newcommand{\dbtilde}[1]{\accentset{\approx}{#1}}

\def\de{\mathrm{d}}

\newcommand{\grrL}{{\dbtilde {R}}}
\newcommand{\grr}{{\tilde {R}}}

\makeatletter
\DeclareRobustCommand{\loplus}{\mathbin{\mathpalette\dog@lsemi{+}}}
\DeclareRobustCommand{\lotimes}{\mathbin{\mathpalette\dog@lsemi{\times}}}
\DeclareRobustCommand{\roplus}{\mathbin{\mathpalette\dog@rsemi{+}}}
\DeclareRobustCommand{\rotimes}{\mathbin{\mathpalette\dog@rsemi{\times}}}

\newcommand{\dog@rsemi}[2]{\dog@semi{#1}{#2}{-90,90}}
\newcommand{\dog@lsemi}[2]{\dog@semi{#1}{#2}{270,90}}
\newcommand{\dog@semi}[3]{
  \begingroup
  \sbox\z@{$\m@th#1#2$}
  \setlength{\unitlength}{\dimexpr\ht\z@+\dp\z@\relax}
  \makebox[\wd\z@]{\raisebox{-\dp\z@}{
    \begin{picture}(1,1)
    \linethickness{\variable@rule{#1}}
    \roundcap
    \put(0.5,0.5){\makebox(0,0){\raisebox{\dp\z@}{$\m@th#1#2$}}}
    \put(0.5,0.5){\arc[#3]{0.5}}
    \end{picture}
  }}
  \endgroup
}
\newcommand{\variable@rule}[1]{
  \fontdimen8  
  \ifx#1\displaystyle\textfont3\else
    \ifx#1\textstyle\textfont3\else
      \ifx#1\scriptstyle\scriptfont3\else
        \scriptscriptfont3\relax
  \fi\fi\fi
}
\makeatother
\begin{document}

\preprint{APS/123-QED}

\title{Logarithmic supertranslations as asymptotic symmetries of gravity at null infinity}

\author{Oscar Fuentealba}
\email{ofuentealba@unap.cl}
\affiliation{Instituto de Ciencias Exactas y Naturales (ICEN), Universidad Arturo Prat, Playa Brava 3256, 1111346 Iquique, Chile}
\affiliation{Facultad de Ciencias, Universidad Arturo Prat, Avenida Arturo Prat Chac\'on 2120, 1110939 Iquique, Chile}
\affiliation{Université Libre de Bruxelles and International Solvay Institutes, ULB-Campus Plaine CP231, B-1050 Brussels, Belgium}

\author{Marc Henneaux}
\email{marc.henneaux@college-de-france.fr}
\affiliation{Université Libre de Bruxelles and International Solvay Institutes, ULB-Campus Plaine CP231, B-1050 Brussels, Belgium}
\affiliation{Collège de France, Université PSL, 11 place Marcelin Berthelot, 75005 Paris, France}

\author{Sébastien Robert}
\email{sebastien.robert@ulb.be}
\affiliation{ Universit\'e Libre de Bruxelles, BLU-ULB Brussels Laboratory of the Universe, C.P. 231, B-1050 Bruxelles, Belgium} 

\author{Céline Zwikel}
\email{celine.zwikel@college-de-france.fr}
\affiliation{Collège de France, Université PSL, 11 place Marcelin Berthelot, 75005 Paris, France}

\date{\today}

\begin{abstract}
\noindent  Logarithmic supertranslations have been shown recently to be symmetries of the gravitational field at spatial infinity.  We extend this work by proving explicitly that they are also symmetries at null infinity. We also show that the logarithmic supertranslation charges match at the ``corner" where spatial infinity and  null infinity meet.  This fully establishes that the asymptotic symmetry group of gravity in the asymptotically flat context contains the logarithmic supertranslations.
\end{abstract}

\maketitle

It was recognized long ago that the asymptotic symmetry algebra of $4$-dimensional gravity in the asymptotically flat context is an infinite-dimensional extension of the Poincar\'e algebra, called the Bondi-Metzner-Sachs (BMS) algebra \cite{Bondi:1962px, Sachs:1962zza}.   In addition to the standard Poincar\'e generators, this algebra contains an infinite number of ``pure supertranslations", which can be described as $\ell\geq2$ modes of a function on the sphere developed in spherical harmonics, the lower   
$\ell=0,1$ modes being the ordinary spacetime translations.

The physical importance of this symmetry algebra was brought out in seminal work carried out more recently, where it was related to the memory effect and to soft theorems constraining scattering amplitudes involving soft gravitons.  This is captured by the ``infrared triangle"(see review \cite{Strominger:2017zoo}).  A striking feature of this remarkable work is the appearance of a ``Goldstone boson" associated with the degeneracy of the gravitational vacuum \cite{Strominger:2017zoo,Compere:2016jwb}.  This boundary field is the potential for the electric part of the shear at infinity.

Through a consistent relaxation of the standard boundary conditions  for the gravitational field at spatial infinity, it was  shown in \cite{Fuentealba:2022xsz}  that the BMS algebra could in fact be further extended by an infinite number of ``logarithmic supertranslations" parametrized, as the pure supertranslations,  by the $\ell\geq2$ modes of a function on the sphere (but with the asociated $\ell=0,1$ modes defining physically trivial gauge redundancies).   The enlarged algebra has been called the 
log BMS algebra. 
This log BMS algebra was further extended in 
\cite{Girelli:2026gbr} by dropping the parity conditions imposed in \cite{Fuentealba:2022xsz}, much in the spirit of \cite{Compere:2011ve}.   We shall stick here, however, to the original parity-restricted log BMS algebra of \cite{Fuentealba:2022xsz}.

The logarithmic supertranslation charges can be shown to be canonically conjugate to the pure supertranslation charges, enabling a supertranslation-free definition of the angular momentum (through nonlinear redefinitions) \cite{Fuentealba:2022xsz,Fuentealba:2023hzq}. The physical significance of the logarithmic supertranslation charges was further explored in \cite{Fuentealba:2023syb}, where it was shown that they coincide at null infinity with the Goldstone mode associated with supertranslation breaking considered in \cite{Strominger:2017zoo} and so did have already a familiar, clear physical interpretation. This construction matches the invariant angular momentum definitions given at null infinity in \cite{Javadinezhad:2018urv,Compere:2019gft,Chen:2021szm,Compere:2021inq,Chen:2021kug,Javadinezhad:2022hhl,Chen:2022fbu,Compere:2023qoa}, providing thereby a  symmetry origin of these developments.

However, even though some partial matching between spatial infinity and null infinity was performed in \cite{Fuentealba:2023syb,Girelli:2026gbr}, the action of the logarithmic supertranslations was formulated only at spatial infinity. What was not achieved was the writing of the logarithmic supertranslations as asymptotic symmetries at null infinity.  It is the purpose of this letter to complete this task by providing a direct, autonomous derivation of the logarithmic supertranslations at null infinity.  We also derive the corresponding charges at null infinity and indicate how they match the charges at spatial infinity.

We focus on future null infinity but the same construction can be done at past null infinity by formally replacing the retarded time $u$ for the advanced time $v$,  $u\to-v$. 

The key step for displaying the logarithmic supertranslations at null infinity resides in a relaxation of the 
 Bondi gauge condition $ g_{rr}=0$ \cite{Bondi:1962px, Sachs:1962zza}, 
which is imposed in the standard treatments.
Here, $r$ is the radial coordinate that reaches null infinity $\mathscr I$ when $r\to\infty$ and $x^A=(\theta,\phi)$ are the coordinates on the celestial sphere.  

To motivate this softening of the gauge conditions, we briefly recall the situation in electromagnetism, where similar logarithmic gauge transformations appear \cite{Fuentealba:2023rvf}.  The analog of the Bondi gauge at null infinity is the condition $A_r = 0$ on the radial component of the vector potential.  However, in order to allow logarithmic gauge transformations with gauge parameter $\epsilon \sim f(x^A) \frac{\log r}{r}$, one must relax this condition as $A_r = o(r^{-1})$ \cite{Fuentealba:2025ekj}. 

Now, by following the logarithmic supertranslations from spatial infinity to null infinity, one finds that these are generated at null infinity by vector fields with component $\xi^u$ behaving as $\xi^u \sim \frac{\log r}{r^2}$ (instead of $\xi^u = \mathcal O(1)$ for supertranslations) (more on this below).  If one performs such a transformation, one clearly destroys the condition $ g_{rr}=0$ since the Lie derivative $\mathcal L_\xi g_{rr} = 2\xi^u_{,r} g_{ur}$ does not vanish but rather behaves asymptotically as $ \frac{\log r}{r^3}$.  This  suggests that in order to accommodate logarithmic supertranslations at null infinity,  one should replace the strict condition $ g_{rr}=0$ by the weaker condition
\begin{equation}\label{grrrelaxation}
 g_{rr}=\frac1{r^3}
 \left(\grr (u,x^A)+\log r\,\grrL  (u,x^A) \right)+ o(r^{-3})
\end{equation}
for some functions $\grr (u,x^A)$ and $\grrL  (u,x^A)$ \cite{TildeR}. 

We thus adopt as line element
\begin{align}\label{asympt Bondi gauge}
\de s^2=& e^{2\beta}\frac{V}{r}\de u^2  -2e^{2\beta}\de u\,\de r   + g_{rr}\de r^2\nonumber \\
& +g_{AB}(\de x^A-U^A\de u)(\de x^B-U^B\de u)\,,
\end{align}
with $g_{rr}$ given by (\ref{grrrelaxation}) and 
\begin{align}
g_{AB}&=r^2q_{AB}+r\,C_{AB}\\\label{asympt Bondi gauge 2}
&+D_{AB}+\frac14q_{AB}C_{CD}C^{CD}+o\left( r^0\right)\,.
\end{align}
The tensor $q_{AB}$ is the 2-sphere metric whose associated covariant derivative is $\nabla$. The traceless shear $C_{AB}$ is related to the news tensor $N_{AB}:=\partial_u C_{AB}$. The tensor $D_{AB}$ is traceless too because 
 we impose the Bondi--Sachs determinant condition $\sqrt{\gamma}=r^2\sqrt{q}$ as in the usual case, where $\gamma\coloneqq\det(\gamma_{ab})$.  We also maintain the strict gauge condition $g_{rA} = 0$.  The functions $\frac{V}{r} + 1$, $\beta$ and the vector $U^A$ are required to vanish as $r \rightarrow \infty$, which, combined with the radial evolution equations imposes
\begin{subequations}\label{expansions of BUV}
\begin{align}\nonumber
\beta&=\frac{\log r}{8r^2}\partial_{u}\grrL - \frac1{8r^2}\left(\frac{1}{4}C^{AB}C_{AB}-\frac1{2}\partial_{u}\grrL-\partial_{u}\grr \right) \nonumber \\&+o(r^{-2})\,, \qquad V=-r+2M(u,x^A)+o(r^{0})\,,  \\ 
U^A
&=- \frac1{2r^2}\nabla_BC^{AB}+\frac{\log r}{r^3}\,\left(-\frac23 \nabla_BD^{AB}+\frac1{12}\partial^A\partial_u \grrL \right)\nonumber\\& +\frac{1}{r^3}U_{3,0}^A(u,x^A)+o(r^{-2})\,, \label{Eq:V} 
\end{align} 
\end{subequations}
where $M$ is the Bondi mass aspect and $U^A_{3,0}$ the angular momentum aspect, which obey dynamical equations in $u$ sourced by $C_{AB}$.

Before proceeding with the analysis, a few comments are in order.  First, we stress that one can always set $\grr (u,x^A)$ and  $\grrL  (u,x^A)$ equal to zero by a logarithmic diffeomorphism $\xi^u \sim f(x^A) \frac{\log r}{r^2}$ completed by appropriate subdominant terms.  If this is done, the asymptotic form of the metric reduces to the Bondi one.  Accordingly, our boundary conditions relax the Bondi gauge by terms induced by logarithmic diffeomorphisms, and only do that.  One might envision combining this relaxation with more drastic relaxations considered in the literature \cite{Geiller:2025dqe,Freidel:2021fxf,Geiller:2022vto,Geiller:2024amx,Geiller:2024ryw,1985FoPh...15..605W,Chrusciel:1993hx, McNees:2024iyu,Compere:2018ylh,Barnich:2011mi,Rignon-Bret:2024gcx,Campoleoni:2023fug,Campiglia:2014yka}.  These interesting extensions fall beyond the scope of our paper. 

Second, one might think that because the extension is by a mere diffeomorphism, it is physically irrelevant.  The point is that  the diffeomorphism that brings one to the Bondi gauge is an improper gauge transformation with non-vanishing charge.  Hence, imposing the stronger condition $g_{rr}=0$ is illegitimate.   It removes physical information rather than just eliminating irrelevant redundancy.

Third, although $g_{rr}$ does not vanish, our boundary conditions are still manifestly compatible with asymptotic flatness because $g_{rr} = o(r^{-2})$. Indeed, taking $\Omega\sim 1/r$ as the conformal factor, the conformally rescaled metric $\Omega^2 g_{\mu\nu}$ matches the conformally rescaled flat metric to leading order near infinity.  Furthermore, since we relax the Bondi gauge by a mere diffeomorphism, the curvature invariants are unchanged and the peeling/non-peeling properties of the Weyl tensor are unaffected.  Note that  the vector field $\frac{\partial}{\partial r}$ is not null because $g_{rr} \not=0$ but only asymptotically so, while the hypersurfaces $u=$ constant are not null because $g^{uu} \not=0$ but only asymptotically so.

Finally, the coefficients appearing in the asymptotic development of the metric, such as $\grr (u,x^A)$ or $\grrL  (u,x^A)$,  involve both parities under the sphere antipodal map.  So, why did we state above that we were considering  logarithmic supertranslations restricted by definite parity conditions?  This is because the parity properties of the fields at spatial infinity are encoded in their asymptotic behaviour at null infinity in the sense that different parities with same asymptotic behaviours at spatial infinity lead to different asymptotic behaviours at null infinity.  Therefore, by reading the asymptotic behaviour at null infinity, one can infer the parity properties at spatial infinity (and incidentally, the matching conditions between future and past null infinities of \cite{Strominger:2017zoo}, as indicated in \cite{Henneaux:2018hdj}).

This key property is derived by integrating the equations of motion with initial data given on an asymptotically flat Cauchy hypersurface all the way to null infinity, a task most easily done in hyperbolic (Beig-Schmidt)  coordinates \cite{Ashtekar:1978zz,Beig:1982ifu}.   It  is well documented in the literature, going back to the earlier works of \cite{Herberthson:1992gcz,Troessaert:2017jcm,Henneaux:2018cst,Henneaux:2018hdj} and fully developed in \cite{Henneaux:2018mgn,Henneaux:2019yqq} and more recently in \cite{Compere:2023qoa,Fuentealba:2024lll,Compere:2025tzr,Fuentealba:2025ekj,Briceno:2025cdu,Compere:2025bnf,Fuentealba:2025ekj,Compere:2026jmk,Compere:2026mdj}.   For that reason, we shall only recall here the salients features of the derivation in the simple case of the scalar field and extend then the reasoning to the parameters of the logarithmic supertranslations, which is relevant to our study.

In four dimensional Minkowski spacetime, one usually takes as asymptotic conditions for a massless Klein-Gordon field the behaviour $\phi \sim \bar \phi/r$  and $\pi \sim \bar \pi /r^2$ on flat spacelike hyperplanes, where $\pi = \dot{\phi}$ is the momentum conjugate to $\phi$.  Here $\bar \phi$ and $\bar \pi$ are functions of the angles (and of time).  Now, the even part of $\bar \phi$ and the odd part of $\bar \pi$ under the sphere antipodal map on the initial surface lead to a solution belonging to the so-called $P$-branch, which, on the hyperboloid of \cite{Ashtekar:1978zz,Beig:1982ifu}, is even under the hyperbolic reflection combining the sphere antipodal map with a hyperbolic time reflection $\tau \rightarrow - \tau$.   At null infinity, the $P$-branch behaves as $1/r$ (and higher order terms in the $1/r$ expansion).  Similarly,  the odd part of $\bar \phi$ and the even part of $\bar \pi$ lead to a solution belonging to the so-called $Q$-branch, which is odd under the hyperbolic reflection. At null infinity it behaves as $\log r/r$ (and subleading terms). It is a general rule that, for identical asymptotic behaviour at spatial infinity, the $Q$-branch dominates the $P$-branch at null infinity (see e.g. appendix of \cite{Henneaux:2018mgn} for precise information).
Thus, if there is a dominant logarithmic term of the form $\log r/r$ in the expansion of $\phi$ near null infinity, one knows that there is an odd part in  $\bar \phi$ and/or an even part in $\bar \pi$ at spatial infinity, while if the expansion starts at $1/r$ at null infinity,  $\bar \phi$ is even and $\bar \pi$ is odd (and can be explicitly related to the two parities of the coefficient of the $1/r$ term at null infinity \cite{Fuentealba:2024lll}).   

Concerning the parameters of the gauge transformations  of electromagnetism, one finds that the $\mathcal O(1)$ angle-dependent $u(1)$ parameter is controlled by the $Q$-branch and behaves as $\mathcal O(1)$ at null infinity, while  the logarithmic angle-dependent $u(1)$ parameter, which is dominant at spatial infinity, has a coefficient that is controlled by the $P$-branch and becomes therefore subdominant at null infinity, giving a term that behaves as $\log r/r$ at null infinity (details in \cite{Fuentealba:2025ekj}).   

Similarly, for gravity, the  $\mathcal O(1)$ term in $\xi^u$, i.e., the supertranslation term,  is related to the $Q$-branch \cite{Troessaert:2017jcm,Fuentealba:2023syb} and, even though subdominant at spatial infinity, becomes dominant at null infinity with respect to the logarithmic supertranlations, controlled by the $P$-branch \cite{footnoteLogPQbranches}.

We now come back to the equations of motion for the Bondi mass aspect and the angular momentum aspect.
As in the standard case, the shear $C_{AB}(u, x^A)$ is a free data and can be usefully decomposed as
$ C_{AB} :=    C_{AB}^{(E)}+C_{AB}^{(B)}\,, \quad  C_{AB}^{(E)} := -2\nabla_{\langle A}\nabla_{B\rangle} C $,
where $C_{AB}^{(E)}$ is its electric part, 
while $C_{AB}^{(B)}$ is its magnetic part.  Here, $\nabla_{\langle A}\nabla_{B\rangle}$ denotes tracefree symmetrization. The ``Goldstone field" $C$ appearing in the expression for $C_{AB}^{(E)}$  \cite{Strominger:2017zoo} has no $\ell=0$ or $\ell=1$ spherical harmonics.

The functions $\grr (u,x^A)$ and $\grrL  (u,x^A)$ accounting for the relaxation of the Bondi gauge are at this stage independent arbitrary functions of $u$.  As the derivation of the charges given below indicates (and in line with the spatial infinity analysis), only one function among these two is physically relevant, the other describing redundancy.  It is useful to define a field that captures the physical content in $\grr (u,x^A)$ and $\grrL  (u,x^A)$, which turns out to be
\begin{equation}\label{tildeC}
    \tilde C(u,x^A)
    :=\frac18 \left(u \partial_u^3\grr +\partial_u^2\grrL \right)\,,
\end{equation}
as the charge analysis shows.   The field $\tilde C(u,x^A)$ can be assumed to have no $\ell=0$ or $\ell=1$ component.  To parallel the expressions for the shear, one can write $\tilde C(u,x^A)$ as $
    \tilde C(u,x^A):=\nabla^{\langle A}\nabla^{B \rangle}\tilde C_{AB}$, with $\tilde C_{AB}$ a pure electric tensor.
We also make the (permissible) choice
\begin{equation}
    \partial_u \tilde C=0 \label{duCtilde}\,.
\end{equation}

The equations of evolution of the Bondi mass aspect and the dressed angular momentum aspect can be written in terms of the Weyl scalars as \cite{footnoteDefinitions}
\begin{align}\label{massloss}
  \partial_uM&=-\frac18N_{AB}N^{AB}+\frac14 \nabla_A \nabla_BN^{AB} \,, \\
    \partial_u \mathcal{P}^L_A&= \nabla^B\mathcal{M}_{AB}+C_{AB}\mathcal{J}^B\label{LorentzEvol}
\end{align}
(see work in preparation \cite{P2026} for the explicit derivation).

Once asymptotic conditions have been adopted, the asymptotic analysis consists in deriving the corresponding asymptotic symmetries and charges.  The computations are conceptually straightforward but somewhat involved.  We will give the results here and make them physically plausible.  The complete derivations will be reported in \cite{P2026}.

We start with the asymptotic symmetries, i.e., with the diffeomorphisms that preserve the asymptotic conditions (\ref{grrrelaxation}) through (\ref{Eq:V}) (residual gauge transformations).  A direct computation shows that these are given by
\begin{subequations}\label{residualsymm}
\begin{align}\nonumber
    \xi^u&=f(u,x^A)\,, \\
    &+ \frac1{r^2}\left(\log r \, \xi^u_{2,1}(u,x^A)+\xi^u_{2,0} (u,x^A)\right)+o(r^{-2})\\
    \xi^r&=  r \, h(x^A) +\mathcal O(r^{0})\,, \;  \,  \, \xi^A=Y^A(x^A)+\mathcal O(r^{-1}) \,,
\end{align}
\end{subequations}
where $Y^A$ are conformal Killing vectors of the $2-$sphere and 
\begin{align}
f&=T(x^A)-u\,h(x^A)\,,\qquad h=-\frac12D_AY^A\,.
\end{align}
The new feature with respect to Bondi gauge is the presence of the subleading terms in $\xi^u$, allowed by our relaxed boundary conditions and  parametrized by the two functions $\xi^u_{2,1}(u,x^A)$ and $\xi^u_{2,0} (u,x^A)$.  There is physical redundancy in this parametrization and only one function, which can be taken to be
\begin{align}
    \tilde T(x^A)&=\frac1{8}\partial_u^2\Big(2\left(4\xi^u_{2,1}-4 \xi^u_{2,0} - u\partial_u\xi^u_{2,1}+2u\partial_u\xi^u_{2,0}  \right) \nonumber
\\ & -2h\grrL -T\partial_u\grr +uh \partial_u\grrL \Big)\,,\label{Eq:TildeT}
\end{align}
is relevant.  The lower harmonics $\ell = 0$ and $\ell = 1$ of $\tilde T$ correspond also to gauge redundancies so that we can assume that $\tilde T(x^A)$ has no $\ell = 0$ or $\ell = 1$ component.

The function  $T(x^A)$ parametrizes the supertranslations (including ordinary spacetime translations) while the function $\tilde T(x^A)$
parametrize the logarithmic supertranslations.  The other independent parameters, contained in $Y^A$, correspond to the homogeneous Lorentz group.

To compute the charges associated with the asymptotic symmetries at null infinity, we use  the covariant phase space formalism and the procedure of  \cite{Iyer:1994ys}. The symplectic current associated with the Einstein--Hilbert Lagrangian is the current $\frac{ \sin \theta}{32\pi G}\delta N_{AB}\,\delta C^{AB}$ of \cite{Ashtekar:1981bq}  supplemented by a boundary term that involves the Goldstone bosons \cite{footnoteOmegaRadial},
\begin{equation}\label{ASsymplecticform}
 \omega^r=\frac{ \sin \theta}{32\pi G}\delta N_{AB}\,\delta C^{AB} +  \frac{ \sin \theta}{32\pi G}\partial_u(8 \, \delta C \,\delta \tilde C)\,.
\end{equation}
The symplectic form vanishes, as it should, in the absence of radiation (taking into account the condition $\partial_u\tilde C=0$). 

The charges are obtained by contracting the symplectic current with a residual symmetry \eqref{residualsymm}
$
I_\xi\omega^r=\partial_u(\delta q_\xi+f_\xi)
$
 upon imposing the equations of motion and up to total derivatives on the sphere. Denoting $\oint$ the integral over a cut of $\mathscr I$, the charge $Q_\xi=\oint q_\xi$ 
is given by
\begin{align}  \label{chargesatscri}
Q_\xi&=\oint\frac{\sin \theta}{16 \pi G}\left( 
T\,q_T+\tilde T\, q_{\tilde T}+ Y^A(q_Y)_A\right)\,,
\end{align}
while $f_\xi[\delta]=\frac{\sin \theta}{32\pi G }\,f\,  N^{AB}\delta \left(C_{AB}-4\tilde C_{AB}\right)$.  Here,
\begin{align} \label{chargesTtilde}
q_T&=4 (M + \tilde C) \,,\qquad 
q_{\tilde T}=-4 C\,, \\
\nonumber
(q_Y)_A&=2\mathcal{P}^L_A-2u\partial_A M  - \frac12C_{AC}\nabla_BC^{BC}\\\label{chargeLorentzscri}
&-\frac{1}{8}\partial_A(
C_{BC}C^{BC})+2(3\tilde C\partial_AC+C \partial_A \tilde C )\,.
\end{align}
Note that if we had allowed the low ($\ell = 0$ and $\ell = 1$) harmonics in $\tilde T$, these  would indeed drop out from the charges as $C$ has no low harmonics.

With respect to the standard case, we have a new charge, the logarithmic supertranslation charge $q_{\tilde T}$, equal to the Goldstone boson $C$ (up to a factor), as found in \cite{Fuentealba:2023syb} from spatial infinity.  The other charges reduce to the  charges of  \cite{Barnich:2011mi} when restricting to the Bondi gauge.   Note that the modification of $M$ by $\tilde C$ does not affect the low harmonics (total mass and linear momentum), in agreement with the findings at spatial infinity \cite{Fuentealba:2022xsz,Girelli:2026gbr}.   

The asymptotic symmetry algebra is given by the bracket of \cite{Barnich:2011mi}, namely $
   \{Q_{\xi_1},Q_{\xi_2}\}_{\text{BT}}:=\delta_{\xi_2}Q_{\xi_1}+F_{\xi_2}[\delta_{\xi_1}]
$, with $F_{\xi}[\delta]=\oint f_\xi[\delta]$.
To carry out this computation, one needs the transformation properties of the fields.  One finds that the field $\tilde C(x^A)$ defined in \eqref{tildeC} transforms as 
$
   \delta \tilde C=  (\mathcal L_Y-3h) \tilde C   + \tilde T(x^A) 
$,
while the covariant mass, dressed angular momentum and the shear transform in the same way as in the Bondi gauge.

We then obtain \cite{P2026}
\begin{align}\label{charge algebra explicite}
  \{Q_{\xi_1},Q_{\xi_2}\}_{\text{BT}}&=Q_{[\xi_1,\xi_2]}+\oint \frac{\sin \theta}{4 \pi G}   (T_1\,\tilde T_2 -\tilde T_1\,T_2)\,.
\end{align}
As $\tilde T$ has only higher-harmonics ($\ell \geq 2$), the central charge is among higher harmonics. The asymptotic symmetry algebra is therefore
\begin{equation}\label{null charge algebra}
    \text{log BMS:}\quad  \mathfrak{iso}(3,1) \loplus \mathfrak{h}(T^{\ell \geq 2},\tilde T^{\ell \geq 2})\,,
\end{equation}
where $\mathfrak{h}(T^{\ell \geq 2},\tilde T^{\ell \geq 2})$ is the Heisenberg algebra generated by the conjugate generators $T^{\ell \geq 2}$ and $\tilde T^{\ell \geq 2}$.  In the limit $u\to \pm \infty$, the flux $f_\xi$ vanishes and the charges form an algebra for the standard bracket, again in perfect agreement with the findings at spatial infinity  \cite{Fuentealba:2022xsz,Girelli:2026gbr}.  This algebra provides a ray representation of the algebra of the asymptotic symmetries (which can be computed directly).  Because of the central charge, one can rewrite it as the direct sum of the Poincar\'e algebra and an infinite number of copies of the Heisenberg algebra \cite{Fuentealba:2022xsz,Girelli:2026gbr,Fuentealba:2023hzq}.

We explained above how the logarithmic diffeomorphisms at spatial infinity and null infinity matched and so defined the same symmetry transformations.  It is then guaranteed that the corresponding charges match in the overlapping regions of definition, i.e., at the ``corners" where the infinities meet.   More precisely,  the charges at future null infinity tend to the charges at spatial infinity in the limit $u \rightarrow - \infty$ and similarly for past null infinity.  
This implies that the conserved charges at spatial infinity will lead to antipodal matching relations. These will be explicitly derived in \cite{P2026}. Also it will be verified there that the log BMS group is a symmetry at timelike infinity, which matches the null infinity description.

To conclude, our results indicate that the natural asymptotic symmetry algebra governing the asymptotic structure of the gravitational field, and in particular gravitational scattering, is the log BMS algebra.    
 The enlargement of the asymptotic symmetry has no impact on the existing soft theorems since the antipodal matching of the Bondi fields are unchanged \cite{P2026}. 
 However, the new symmetry does imply new Ward identities and potential new memory effect.  We are currently exploring their consequences.

It has been argued recently that the logarithmic supertranslations play an essential role in a proper description of dressed states
\cite{Bosma:2026qcb}.   Our work puts that statement on a firmer basis.

\begin{acknowledgments}

The work of MH was partially supported by FNRS-Belgium (convention IISN 4.4503.15), as well as by research funds from the Solvay Family.  O.F. is grateful to the Collège de France and the Université Libre de Bruxelles for kind hospitality
while this work was completed. This research has been partially supported by ANID through the Fondecyt grant N° 11251195. CZ is grateful for discussions with Mathieu Beauvillain, Blagoje Oblak, Romain Ruzziconi and Simone Speziale.  SR thanks Geoffrey Compère for collaboration on related topics. 
\end{acknowledgments}



\begin{thebibliography}{99}
\bibitem{TildeR}
 We take independent functions $\grr (u,x^A)$ and $\grrL  (u,x^A)$ because once diffeomorphisms of the form $\xi^u \sim \frac{\log r}{r^2}$ are allowed, it is convenient to allow also subdominant  terms of the form $\xi^u \sim \frac{1}{r^2}$ which modify $\grr (u,x^A)$ without changing $\grrL  (u,x^A)$.  This freedom corresponds to a proper gauge one with no new independent charge (see (\ref{chargesatscri}) and (\ref{Eq:TildeT})).
\bibitem{footnoteDefinitions}
The quantities appearing in (\ref{massloss}) and (\ref{LorentzEvol}) are defined by
\begin{subequations}
\begin{align}
 \mathcal P^{A}_L &= \mathcal P^A+\frac18 \nabla^A\partial_u\left( \grr +\frac56 \grrL\right) \nonumber \\
\mathcal{P}_A &=-\frac{3}{2}U^3_A+\frac{3}{32}\partial_A(C_{BC}C^{BC}) \nonumber \\&+\frac{3}{4}C_{AB}\nabla_CC^{BC}-\frac{4}{3}\nabla^BD_{AB}, \nonumber\\
\mathcal M_{AB}&=\mathcal M\,q_{AB}+\widetilde{\mathcal M}\,\epsilon_{AB} \nonumber\\
\mathcal M&= M+\frac{1}{16}\partial_u(C_{AB}C^{AB}), \nonumber\\
\widetilde{\mathcal M}&=\frac{1}{4}\left(\nabla_A\nabla_B-\frac{1}{2}N_{AB}\right)\widetilde{C}^{AB},\nonumber
\\
\mathcal J_A&=\frac{1}{2}\nabla^BN_{AB},\nonumber\\
\mathcal N_{AB}&=\frac{1}{2}\partial_uN_{AB}.\nonumber
\end{align}
\end{subequations}
The quantity $ \mathcal P^{A}_L$ can be viewed as a dressing by the logarithmic functions $\grr (u,x^A)$ and $\grrL  (u,x^A)$  of the covariant angular momentum  $\mathcal P_A$ defined in the Bondi gauge.
\bibitem{footnoteOmegaRadial}
For our purposes, only the radial component of the symplectic current is relevant, so we restrict our discussion to this component.
\end{thebibliography}
\end{document}